\documentclass[aps,twocolumn,groupedaddress,showpacs]{revtex4-1}
\usepackage{epstopdf}
\usepackage{physics}
\usepackage{tikz,lipsum,lmodern}
\usepackage[export]{adjustbox}
\usepackage{amsmath}
\usepackage{amssymb}
\usepackage{svg}

\usepackage[unicode=true,pdfusetitle,
 bookmarks=true,bookmarksnumbered=false,bookmarksopen=false,
 breaklinks=false,pdfborder={0 0 0},backref=false,colorlinks=true,citecolor=blue,urlcolor=violet 
]{hyperref}
\usepackage{xcolor}
\usepackage{graphicx}
\usepackage{setspace}
\begin{document}
\title{Dynamics of Quantum Coherence and Non-Classical Correlations in Open Quantum System Coupled to a Squeezed Thermal Bath}

\author{Neha Pathania}
\email{neha@ctp-jamia.res.in}
\author{Ramniwas Meena}
\email{meena.53@iitj.ac.in}
\author{Subhashish Banerjee }
\email{subhashish@iitj.ac.in}

\affiliation{$^1$Department of Physics, Indian Institute of Technology, Jodhpur, Rajasthan-342030, India}

\begin{abstract}
We investigate the intricate dynamics of quantum coherence and non-classical correlations in a two-qubit open quantum system coupled to a squeezed thermal reservoir. By exploring the correlations between spatially separated qubits, we unravel the complex interplay between quantum correlations and decoherence induced by the reservoir. Our findings demonstrate that non-classical correlations such as quantum consonance, quantum discord, local quantum uncertainty, and quantum Fisher information are highly sensitive to the collective regime. These insights identify key parameters for optimizing quantum metrology and parameter estimation in systems exposed to environmental interactions. Furthermore, we quantify these quantum correlations in the context of practical applications such as quantum teleportation,  using the two metrics \textit{viz.} maximal teleportation fidelity and fidelity deviation. This work bridges theoretical advancements with real-world applications, offering a comprehensive framework for leveraging quantum resources under the influence of environmental decoherence.


\end{abstract}

\maketitle

\section{Introduction}

The study of non-classical correlations in quantum systems remains a fundamental challenge in quantum information processing and a crucial test for quantum theory. Various theoretical and experimental methods for creating non-classical correlations have been extensively explored in the literature~\cite{loudon1980non,riedinger2016non,xu2010experimental,ur2023generating,businger2022non,adesso2016measures}. However, it is well established that the correlations of quantum systems are significantly influenced by their interaction with the environment, as these systems inevitably exist within an open quantum framework~\cite{breuer2002theory,banerjee2018open,maziero2009classical,slaoui2018universal,tan2016unified}. To mitigate dissipation and decoherence arising from environmental interactions, substantial advancements have been achieved, particularly in developing strategies to counteract unavoidable noise sources, such as environmental coupling. Interestingly, recent insights reveal that noise and environmental interactions are not always detrimental and can, under specific conditions, be leveraged to achieve beneficial outcomes ~\cite{plenio1999cavity,brask2015autonomous,braun2002creation,benatti2003environment,kim2002entanglement,srikanth2007environment}.

While many approaches focus on transient entanglement, it has been demonstrated that steady-state entanglement can emerge from dissipative dynamics ~\cite{kraus2008preparation,verstraete2009quantum,diehl2008quantum}. Experimental and theoretical studies have investigated steady-state entanglement in various platforms, including trapped ions \cite{schneider2002entanglement}, cavity-confined atoms ~\cite{kastoryano2011dissipative}, superconducting qubits \cite{reiter2013steady}, spin systems ~\cite{schuetz2013steady}, and nanomechanical systems \cite{walter2013entanglement}, with notable experimental validations ~\cite{shankar2013autonomously,barreiro2011open,krauter2011entanglement,lin2013dissipative}. These methods often rely on engineered dissipation and quantum bath engineering ~\cite{reiter2012driving,aron2014steady,vacanti2009cooling}, where the system is driven toward a unique fixed point that corresponds to an entangled state. Crucially, these techniques require an external coherent driving field, effectively serving as a source of work.

A critical aspect of this research is understanding how quantum correlations are impacted by noise, which manifests as an open system effect ~\cite{dhar2013controllable,naikoo2019facets,banerjee2017characterization}. Recent, experimental studies have explored entanglement dynamics using continuous environmental monitoring through quantum trajectory techniques \cite{strunz1999open,nha2004entanglement,daley2014quantum}. In this work, we investigate the influence of noise on the entanglement generated between two spatially separated qubits through their interaction with a bath initially in a squeezed-thermal state. This analysis is particularly relevant for assessing the performance of two-qubit gates in practical quantum information systems. Starting from an uncorrelated state, the qubits become entangled over time via their interaction with the bath, but the entanglement is eventually degraded due to the quantum-to-classical transition induced by noise.

In the study of entanglement within two-qubit systems, which rapidly evolve into mixed states, measures suitable for mixed state entanglement (MSE) are crucial. While entanglement in pure bipartite states is well-defined, MSE lacks a universal definition. Existing approaches, such as entanglement of formation ~\cite{wootters1998entanglement,bennett1996mixed} and separability \cite{werner1989quantum}, highlight the complexity of capturing MSE with a single metric. Prior studies have introduced frameworks like a probabilistic approach, where MSE is represented by a probability density function (PDF) \cite{bhardwaj2008characteristics}, with metrics such as concurrence and negativity emerging as specific components of this framework \cite{banerjee2010entanglement,banerjee2010dynamics}. Building on these advancements, additional correlation and coherence measures are essential to capture aspects of quantum systems beyond entanglement. They provide a more complete understanding of system dynamics, validate findings, and reveal insights into quantum resources critical for practical applications.

The characterization of non-classical correlations along with quantum coherence is a central challenge in understanding open quantum systems, particularly under environmental interactions ~\cite{chandrashekar2007symmetries}. Measures such as the relative entropy of coherence~\cite{baumgratz2014quantifying,song2020evolution} quantify a system’s superposition properties, while tools like concurrence~\cite{wootters1998entanglement,wootters2001entanglement,rungta2001universal}, quantum discord~\cite{ollivier2001,chen2011quantum,ali2010quantum}, and quantum consonance~\cite{pei2012using} capture different aspects of quantum correlations. Metrics like local quantum uncertainty (LQU)~\cite{luo2003wigner,he2014measuring,PhysRevLett.110.240402} and quantum Fisher information (QFI) ~\cite{liu2020quantum,abdel2012quantum,zidan2019quantum} further illuminate non-classical correlations and their sensitivity to environmental effects, especially in non-Markovian regimes. In a two-qubit system, local measurements are made on each qubit, and LQU measures the uncertainty associated with the local outcomes~\cite{PhysRevLett.110.240402}. It quantifies how much of a correlation is localized in one part of the composite system. Higher LQU implies that local measurements significantly disturbs the system. Studying Quantum Fisher Information (QFI) in a two-qubit system ~\cite{liu2010quantum} plays a crucial role in understanding and optimizing quantum metrology and parameter estimation within quantum systems that are subject to environmental interactions. In a two-qubit open quantum system, QFI is used to evaluate how effectively a parameter influencing the dynamics ( e.g. interaction strength, decoherence rate) can be estimated. These measures reveal how memory effects can enhance or degrade quantum resources, impacting coherence, correlations, and applications like quantum communication. In particular, teleportation fidelity and fidelity deviation serve as benchmarks for assessing the stability of quantum information transfer in noisy environments~\cite{liu2020experimental,ghosal2020fidelity,ghosal2021characterizing}. Earlier studies have shown that the mixed entangled states can be useful for quantum teleportation~\cite{bandyopadhyay2006quantum,chakrabarty2011studyquantumcorrelationsopen}.\\
In this work, for an open quantum system consisting of 2-qubits and interacting with a squeezed thermal bath, we consider two scenarios: (i) the collective regime, where the two qubits are close enough to experience collective decoherence due to shared interaction with the bath and (ii) when the bath exhibits a long correlation length, or the independent regime, where the qubits are spatially separated. Here, we examine the dynamics of coherence and correlations, utilizing these measures to explore their evolution, practical utility, and implications for robust quantum information processing and communication.\\
This work is structured as follows. Section \textcolor{purple}{II} provides the theoretical background, introducing key quantifiers of coherence, quantum correlations, and information-theoretic measures such as local quantum uncertainty (LQU) and quantum Fisher information (QFI). Additionally, we discuss their practical relevance through metrics like teleportation fidelity and fidelity deviation. Section \textcolor{purple}{III} details the system models under consideration, including the collective and independent regimes of bath interactions. In Section \textcolor{purple}{IV}, we analyze the temporal evolution of coherence and correlations as functions of system parameters such as interqubit distance and environmental characteristics. Finally, Section \textcolor{purple}{V} summarizes the findings and their implications for robust quantum information processing in open quantum systems.
\section{Preliminaries}
In this section, we defined the measures of coherence and non-classical correlations and teleportation metrics.
\subsection{Relative Entropy of Coherence}
Quantum coherence is inherently fragile, and its degradation due to environmental noise in open quantum systems, commonly termed decoherence, is a hallmark of the quantum-to-classical transition that becomes evident at macroscopic scales ~\cite{breuer2002theory,zurek2003decoherence, schlosshauer2004decoherence}. Here, we will study the dynamical evolution of a coherence measure named the relative entropy of coherence (~\cite{baumgratz2014quantifying,streltsov2017colloquium}) in a non-Markovian system. Among the different coherence measures, the relative entropy of coherence holds an operational significance within the framework of resource theory ~\cite{winter2016operational}. For a density matrix \(\rho\), it is defined as~\cite{baumgratz2014quantifying}:
\begin{equation}
    C_{\text{rel}}(\rho) = S(\rho_{\text{diag}}) - S(\rho),
\end{equation}
where, $
S(\rho) = -\text{Tr}(\rho \log \rho)
$ is the von Neumann entropy of the state $\rho$, and $\rho_{\text{diag}}$ is the diagonal part of $\rho$ in the reference basis. This measure satisfies desirable properties such as non-negativity, monotonicity (under CPTP operations), and additivity ~\cite{baumgratz2014quantifying}. In the context of two-qubit systems that interact with non-Markovian environments, such as a squeezed thermal bath, the relative entropy of coherence provides valuable insights into quantum coherence dynamics ~\cite{rivas2014quantum}.
\subsection{Concurrence}
Concurrence is a key measure of quantum entanglement, quantifying correlations in a quantum state. Introduced for two-qubit systems, it links directly to the entanglement of formation, reflecting the resources needed to create the entangled state ~\cite{wootters1998entanglement,wootters2001entanglement}.

\begin{equation}
C_E(\rho) = \max\{0, \sqrt{ \lambda_1} - \sqrt{ \lambda_2 }- \sqrt{\lambda_3} - \sqrt{\lambda_4}\},
\end{equation}
where $\lambda_i's $ are the eigenvalues of the matrix $\rho \tilde{\rho}$ in descending order, with $\tilde{\rho} = (\sigma_y \otimes \sigma_y) \rho^* (\sigma_y \otimes \sigma_y)$. Here, $\rho^*$ is the complex conjugate of $\rho$, and $\sigma_y$ is the Pauli-Y matrix. This framework has been generalized to higher-dimensional systems, offering a powerful approach to study entanglement in arbitrary dimensions through the universal state inversion method ~\cite{rungta2001universal}. For pure states, concurrence simplifies to a closed-form expression, making it highly versatile for both theoretical analysis and experimental implementations. Importantly, concurrence remains an essential tool in investigating the dynamics of entanglement in system-environment interactions, such as entanglement decay and revival in non-Markovian open quantum systems~\cite{maziero2010,yu2009}.

\subsection{Quantum Discord}
 Quantum discord has emerged as a pivotal measure for quantifying quantum correlations, extending beyond entanglement to capture non-classical features in separable states. This makes it particularly valuable in open quantum systems, where it provides insights into non-Markovian dynamics, system-environment interactions, and decoherence effects~\cite{modi2012,adhikari2012operational}. Quantum discord has been linked to quantum advantages in areas such as metrology, thermodynamics, and decoherence-free subspaces, and plays a significant role in studying entanglement-independent correlations in quantum computation and communication~\cite{ollivier2001quantum}.  

For a two-qubit state $\varrho$, quantum discord quantifies the difference between quantum and classical mutual information:  
\begin{equation}
\label{eq:discord_def}
\mathcal{Q_D}(\varrho) = \mathcal{I}(\varrho) - \mathcal{C}(\varrho),
\end{equation}  
where $\mathcal{I}(\varrho)$, the quantum mutual information, measures total correlations, and $\mathcal{C}(\varrho)$, the classical mutual information, quantifies correlations accessible through local measurements. \\ 
Calculating quantum discord is computationally intensive due to the optimization over measurements, but analytical expressions exist for certain states, such as two-qubit X-states~\cite{luo2008quantum,ali2010quantum,chen2011quantum,dahbi2023intrinsic}. The density matrix of an X-state in the computational basis has the form:  
\begin{equation}
\label{eq:x_state}
\varrho = 
\begin{pmatrix}
\varrho_{11} & 0 & 0 & \varrho_{14} \\
0 & \varrho_{22} & \varrho_{23} & 0 \\
0 & \varrho_{32} & \varrho_{33} & 0 \\
\varrho_{41} & 0 & 0 & \varrho_{44}
\end{pmatrix}.
\end{equation}  
For such states, the quantum mutual information is given as,
\begin{equation}
    \mathcal{I}(\varrho) = \mathcal{S}(\varrho_A) + \mathcal{S}(\varrho_B) + \sum_{j=0}^3 \lambda_j log_{2} \lambda_j ,
\end{equation}
where, \(\varrho_A\) and \(\varrho_B\) are the marginal states of \(\varrho\) and \(\lambda_j\)'s are the eigenvalues of the density matrix 
 \(\varrho\). The simplified expressions of \(\lambda_j\)'s and von-Neumann entropy \(\mathcal{S}(\varrho_A)\), \(\mathcal{S}(\varrho_B)\) are given in ~\cite{ali2010quantum}.
 The classical mutual information is obtained by optimizing the conditional entropy over local measurements on one subsystem:  
\begin{equation}
\label{eq:classical_info}
\mathcal{C}(\varrho) = \mathcal{S}(\varrho_A) - \min_{\{\Pi_B^i\}} \sum_i p_i \mathcal{S}(\varrho_{A|i}),
\end{equation}  
where $\{\Pi_B^i\}$ are projective measurements on $B$, $p_i$ is the probability of outcome $i$, and $\varrho_{A|i}$ is the post-measurement state of $A$. To calculate the classical mutual information and consequently quantum discord, we have to minimize the quantity \(\mathcal{S}(\varrho_{A|i})\) w.r.t the von- Neumann measurements. This is given in detail in ~\cite{ali2010quantum,dahbi2023intrinsic}.

\subsection{Quantum Consonance}
Quantum correlations arising from quantum consonance have garnered significant attention in recent studies ~\cite{pei2012using,motavallibashi2021non,elghaayda2022local,khedif2021thermal}. Quantum consonance offers a way to quantify non-classical correlations beyond entanglement, and it is derived by isolating the local coherence from the total coherence using local unitary operations. This measure effectively characterizes the sum of entanglement and other quantum correlations ~\cite{pei2012using}. Mathematically, quantum consonance is expressed as:
\begin{align}
\label{eq:qc_def}
\mathcal{Q_C}(\rho) = \sum_{ijmn} \left| \rho_{ijmn}^{c}(1-\delta_{im})(1-\delta_{jn}) \right|,
\end{align}
where $\rho^c = (U_1 \otimes U_2) \rho_{AB} (U_1 \otimes U_2)^{\dagger}$ represents the modified state obtained by applying specific local unitary operations $U_1$ and $U_2$ on the original density matrix $\rho$. The Kronecker delta $\delta_{im}$ ensures that only off-diagonal terms are considered. 

For two-qubit X-states, as defined in Eq.~\eqref{eq:x_state}, quantum consonance simplifies significantly due to the structure of the density matrix. In this case, quantum consonance reduces to:
\begin{equation}
\label{eq:qc_xstate}
\mathcal{Q_C}(\rho) = 2\left( |\varrho_{23}| + |\varrho_{14}| \right).
\end{equation}

This simplification highlights that for X-states, quantum consonance is directly proportional to the magnitude of the off-diagonal terms in the density matrix. Such a property makes it a useful tool for quantifying quantum correlations in systems where the structure of the density matrix allows clear separation of contributions from coherence and entanglement. Quantum consonance has applications in analyzing quantum thermodynamic systems, quantum channels, and the behavior of correlated subsystems in non-Markovian environments ~\cite{khedif2021thermal, motavallibashi2021non}.

\subsection{Local Quantum Uncertainity}
Local quantum uncertainty (LQU) is a measure of quantum correlations in a bipartite quantum system. Introduced as a geometric quantifier, LQU captures the minimal disturbance to one subsystem caused by a local measurement on the other subsystem. It is based on the skew information, which quantifies the incompatibility between a quantum state and an observable. Unlike entanglement measures, LQU is nonzero even in separable states, making it a broader indicator of quantum correlations.
 For a two-qubit system, the LQU can be computed analytically when the state $\rho_{AB}$ is in the X-state form ~\cite{PhysRevLett.110.240402}:
\begin{equation}
    \mathcal{U}(\rho) = \min_{H_\Lambda} \mathcal{I}(\rho, H_\Lambda),\label{eq:lqu_def}
\end{equation}

where \( H_\Lambda = H_A \otimes I_B \) represents a local observable acting on subsystem \( A \), parameterized by a non-degenerate spectrum \(\Lambda\), and \( \mathcal{I}(\rho, H_\Lambda) \) is the skew information. The skew information, which measures the non-commutativity between the quantum state \(\rho\) and the operator \(H_\Lambda\), is given by~\cite{wigner1963information,luo2003wigner}:
\begin{equation}
\mathcal{I}(\rho, H_\Lambda) = -\frac{1}{2} \mathrm{Tr} \left([\rho^{1/2}, H_\Lambda]^2 \right).
\label{eq:skew_info}
\end{equation}

For a bipartite system of dimensions \( 2 \otimes d \), the LQU simplifies to a closed form~\cite{PhysRevLett.110.240402,wu2018local}:
\begin{equation}
\mathcal{U}(\rho) = 1 - \lambda_{\mathrm{max}}(W_{AB}),
\label{eq:lqu_closed_form}
\end{equation}
where \( \lambda_{\mathrm{max}}(W_{AB}) \) is the largest eigenvalue of the \(3 \times 3\) symmetric matrix \(W_{AB}\). The matrix elements of \(W_{AB}\) are defined as:
\begin{equation}
(W_{AB})_{ij} = \mathrm{Tr} \left[\rho^{1/2} (\sigma_i \otimes I_B) \rho^{1/2} (\sigma_j \otimes I_B) \right],
\label{eq:wab_matrix}
\end{equation}
with \(i, j = x, y, z\), and \(\sigma_i\) are the Pauli matrices acting on subsystem \( A \).
The LQU is bounded between 0 and 1, with a higher value indicating stronger quantum correlations.
The LQU provides an efficient way to quantify quantum correlations, using the spectral properties of the symmetric matrix \(W_{AB}\) to avoid the need for direct minimization in higher-dimensional systems.

\subsection{Quantum Fisher Information}
Quantum Fisher Information (QFI) is a fundamental concept in quantum metrology, used to quantify the sensitivity of a quantum state to variations in a parameter \( \vartheta \)~\cite{liu2020quantum}. In this work, QFI is defined based on the symmetric logarithmic derivative (SLD) definition, 
\begin{equation}
    \mathcal{F}_{\mathcal{I}} (\rho_{\vartheta}) =\Tr{\rho_{\vartheta}L_{\vartheta}^2},
\end{equation}
where, the SLD \(L_{\vartheta}\) is given by \(2\partial_{\vartheta} \rho_{\theta}= \rho_{\vartheta}L_{\vartheta} + L_{\vartheta}\rho_{\vartheta}\).
For a quantum state \(\rho_{\vartheta}\), the QFI $\mathcal{F}_{\mathcal{I}}$ for the estimated parameter \(\vartheta\) is given as follows~\cite{wu2018local,elghaayda2022local};
\begin{equation}
    \mathcal{F}_{\mathcal{I}}(\rho_{\vartheta}) = \sum_{m} \frac{(\partial_{\vartheta} \lambda_{m})^2}{\lambda_{m}} + \sum_{m \neq n} \frac{2(\lambda_m - \lambda_n)^2}{\lambda_m + \lambda_n} \big| \langle \Phi_m \big| \partial_{\vartheta}\Phi_n\rangle \big|^2 ,
    \label{fisher}
    \end{equation}
where, \(\lambda_{m}\) is the eigenvalue of the estimated state \(\rho_{\vartheta}\), \(\big|\Phi_m\rangle\) is the corresponding eigenvector and \(\partial_{\vartheta}(.)\) means the partial derivative. \\
For a non-full rank density matrix, the expression of the QFI \(\mathcal{F}_{\mathcal{I}}\) can be rewritten as ~\cite{wu2018local},
\begin{align}
    \mathcal{F}_{\mathcal{I}} (\rho_{\vartheta}) &= \sum_{m=1}^{p} \frac{(\partial_{\vartheta} \lambda_{m})^2}{\lambda_{m}} + \sum_{m=1}^{p} 4 \lambda_{m} \langle \partial_{\vartheta} \Phi_m \big| \partial_{\vartheta} \Phi_m\rangle \nonumber \\
     & - \sum_{m,n= 1}^{p} \frac{8 \lambda_{m}\lambda_{n}}{\lambda_{m}+\lambda_{n}} \big| \langle \Phi_m \big| \partial_{\vartheta}\Phi_{n} \rangle \big|^2 ,
\end{align}
where p is the rank of the density matrix. In the model discussed in this article, the density matrix possesses full rank. In the case of pure state \(\big|\Phi \rangle \), the QFI can be simplified as, 
\begin{equation}
   \mathcal{F}_{\mathcal{I}}(\big|\Phi \rangle) = 4 (\langle \partial_{\vartheta} \Phi \big| \langle \partial_{\vartheta} \Phi \rangle - \big| \langle \Phi \big| \partial_{\vartheta} \Phi \rangle \big|^2 ).
\end{equation} 
\subsection{Teleportation Fidelity and Fidelity Deviation}
Quantum teleportation (QT) is a foundational protocol in quantum communication, enabling the transmission of a quantum state between two parties using shared entanglement and classical communication~\cite{bennett1993teleporting}. Teleportation fidelity \( F \), quantifies the overlap between the input state \( |\psi_{\text{in}}\rangle \) and the teleported output state \( \rho_{\text{out}} \), expressed as~\cite{horodecki1996teleportation}:  
\begin{equation}
    F = \langle \psi_{\text{in}} | \rho_{\text{out}} | \psi_{\text{in}} \rangle. 
\end{equation}  
Perfect QT is achieved when the resource state is maximally entangled, ensuring the output state matches the input exactly. However, practical scenarios involve mixed entangled states, leading to imperfect teleportation.  

The average teleportation fidelity for a two-qubit state \( \rho \) is defined as:  
\begin{equation}
    \langle f_\rho \rangle = \int f_{\psi,\rho} d\psi, 
\end{equation}  
where \( f_{\psi,\rho} = \langle \psi | \chi | \psi \rangle \) denotes the fidelity between the input-output pair \( (|\psi\rangle \langle \psi |, \chi) \). A two-qubit state is considered useful for QT if \( \langle f_\rho \rangle \geq 2/3 \), as \( 2/3 \) represents the classical limit~\cite{horodecki1996teleportation}.  

The maximal teleportation fidelity, \( \mathbf{F}_\rho \), corresponds to the highest achievable average fidelity under local unitary (LU) operations~\cite{horodecki1996teleportation,badziag2000local}:
\begin{equation}
    \mathbf{F}_\rho = \frac{1}{2} \left( 1 + \frac{1}{3} \sum_{i=1}^3 |t_{ii}| \right), 
    \label{maxfidelity}
\end{equation}  
where \( t_{ii} \) are the diagonal elements of the correlation matrix \( T \), defined as \( T_{ij} = \text{Tr}(\rho \, \sigma_i \otimes \sigma_j) \), with \( \sigma_i \) denoting the Pauli matrices~\cite{ghosal2020fidelity}.  

To assess the robustness of QT, fidelity deviation, \( \Delta \mathbf{F}_\rho \), is introduced as the standard deviation of the fidelity over all input states:
\begin{equation}
    \Delta \mathbf{F}_\rho = \sqrt{\langle f_\rho^2 \rangle - \langle f_\rho \rangle^2}, 
\end{equation}  
where \( \langle f_\rho^2 \rangle \) represents the average of the squared fidelity. Its exact form is given by~\cite{ghosal2020fidelity}:
\begin{equation}
    \Delta \mathbf{F}_\rho = \frac{1}{3 \sqrt{10}} \sqrt{ \sum_{i<j = 1}^3 (|t_{ii}| -  |t_{jj}|)^2}. 
    \label{fidelitydev}
\end{equation}  
Here, det T $ < 0 $. A nonzero \( \Delta \mathbf{F}_\rho \) indicates fidelity variations across input states, highlighting the protocol's instability. Thus, both \( \mathbf{F}_\rho \) and \( \Delta \mathbf{F}_\rho \) together provide a comprehensive evaluation of a quantum state’s effectiveness and stability for QT. 

\begin{figure*}
    \centering
    \includegraphics[width=0.48\linewidth]{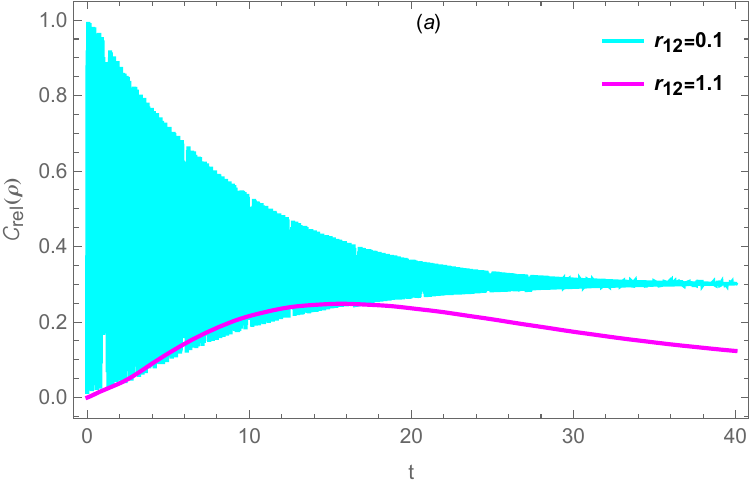}
    \includegraphics[width=0.48\linewidth]{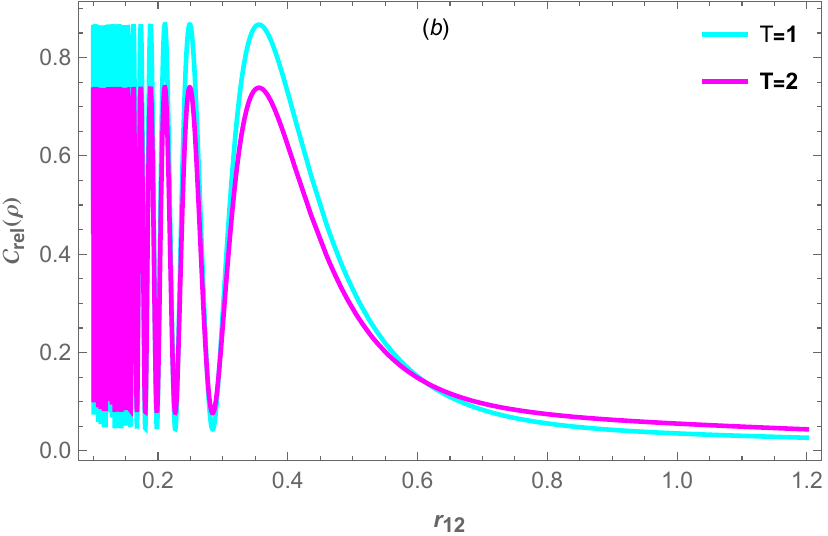}
    \caption{Variation of the Relative Entropy of Coherence, \(C_{rel}(\rho)\)  as a function of a) time (\(t\)) and b) inter-qubit distance (\(r_{12}\)) in a two-qubit system. In a) for the time evolution of \(C_{rel}(\rho)\), the blue curve shows the coherence dynamics in the collective regime (\(r_{12} = 0.1\)), while the pink curve represents the independent regime (\(r_{12} = 1.1\)), with parameters \(T = 1\) and squeezing parameter \(r = 0.35\). In b) \(C_{rel}(\rho)\) is analyzed as a function of \(r_{12}\) at two temperatures, \(T = 1\)(blue line) and \(T = 2\)(pink line), with \(t = 1\) and \(r = 0.35\).
}
    \label{fig:relative_entropy}
\end{figure*}

\section{ Two-Qubit system interacting with a squeezed thermal bath}

We now briefly examine the dynamics of a two-qubit system interacting with a squeezed thermal bath through a dissipative interaction. 
The dissipative interaction between 2-qubits (a two-level atomic system) and the bath (represented as a three-dimensional electromagnetic field (EMF)) via the dipole interaction ~\cite{Agarwal_2012,ficek2002entangled} is described by the Hamiltonian,
\begin{align}
H_{\text{total}} &= H_S + H_B + H_{SB} \nonumber \\
                 &= \sum_{j=1}^2 \hbar \omega_j S_j^{z} \nonumber 
                 + \sum_{\vec{k}s} \hbar \omega_k \left( \alpha_{\vec{k}s}^\dagger \alpha_{\vec{k}s} + \frac{1}{2} \right) \nonumber \\
                 &\quad - i\hbar \sum_{\vec{k}s} \sum_{j=1}^{2} [ \vec{\mu}_j \cdot \vec{g}_{\vec{ks}} \left( \vec{r}_j \right)(S_{j}^{+} + S_{j}^{-}) \alpha_{\vec{k}s} - \text{h.c.} ].
\label{eq:total_hamiltonian}
\end{align}
Here, the two-qubits are modelled as two-level systems with excited state $\big|e_{j} \rangle$, ground state $\big|g_{j} \rangle$, transition frequency $\omega_{j}$, and transition dipole moments $\Vec{\mu_{j}}$. We assume that the qubits are located at different atomic positions $\Vec{r_{j}}$. The transition dipole moments are dependent on $\Vec{r_{j}}$.\\
$S_{j}^+ = \big|e_{j} \big\rangle \big\langle g_{j}\big|$ and $S_{j}^- = \big|g_{j} \big\rangle \big\langle e_{j}\big|$ are the dipole raising and lowering operators, respectively, satisfying the commutation relation and 
\begin{equation}
    S_{j}^{z} = \tfrac{1}{2} (\big|e_{j} \big\rangle \big\langle e_{j}\big| - \big|g_{j} \big\rangle \big\langle g_{j}\big|),
\end{equation}
is the energy operator of the $j^{th}$-atom. 
$\alpha_{\vec{k}s}^\dagger$, $ \alpha_{\vec{k}s}$ are the creation and annihilation operators of the field mode $\vec{k}s$ with wave vector $\vec{k}$, frequency $\omega_{k}$ and polarization index $s = 1, 2$. The system-reservoir coupling constant is expressed as:
\begin{equation}
    \Vec{g}_{\vec{k}s} (\vec{r}_j) = \left(\frac{\omega_{k} }{2\epsilon_0 \hbar V}\right)^{1/2} \Vec{e}_{\vec{k}s} e^{i \vec{k}.\vec{r}_{j}},
\end{equation}
$V$ is the normalization volume, $\epsilon_0$ is the permittivity of free space, and $\Vec{e}_{\vec{k}s}$ is the unit polarization of the field. This equation implies that the system-reservoir coupling constant depends on the atomic position $\vec r_{j}$. 
Assuming separable initial conditions, and after tracing over the bath, the reduced density matrix of the qubit system is derived in the interaction picture under the standard Born–Markov approximation, incorporating the rotating wave approximation (RWA) as detailed in ~\cite{banerjee2010dynamics}
\begin{align}
\dv{\rho}{t} =& - \frac{\iota}{\hbar} [H_{\Tilde{S}}, \rho] \nonumber\\
-& \frac{1}{2} \sum_{m,n=1}^{2} \Gamma_{mn} [1+ \Tilde{N}](\rho S_{m}^+ S_{n}^- +S_{m}^+ S_{n}^- \rho - 2 S_{n}^- \rho S_{m}^+) \nonumber\\
-& \frac{1}{2} \sum_{m,n=1}^{2} \Gamma_{mn} \Tilde{N}( \rho S_{m}^- S_{n}^+ + S_{m}^- S_{n}^+ \rho - 2 S_{n}^+ \rho S_{m}^- \nonumber\\
+& \frac{1}{2} \sum_{m,n=1}^{2} \Gamma_{mn} \Tilde{M}( \rho S_{m}^+ S_{n}^+ + S_{m}^+ S_{n}^+ \rho - 2 S_{n}^+ \rho S_{m}^+ \nonumber\\
+& \frac{1}{2} \sum_{m,n=1}^{2} \Gamma_{mn} \Tilde{M}^* (\rho S_{m}^{-} S_{n}^- + S_{m}^- S_{n}^- \rho - 2 S_{n}^- \rho S_{m}^-) .
\label{eq:master_equation}
\end{align}
In this Eq. \ref{eq:master_equation}, $\Tilde{N} = N_{th}( \cosh{2r}) + \sinh^2{r} $ and $ \Tilde{M}= - \frac{1}{2} \sinh{(2r)} e ^{\iota \varphi} ( 2 N_{th} +1 ) \equiv R e ^{\iota \varphi(\omega_{0})}$ with $\omega_{0} = \frac{\omega_{1}+\omega_{2}}{2}$ and $N_{th} = (e^{\frac{\hbar \omega}{k_B T}}-1)^{-1}$. Here, $N_{th}$ is the Planck distribution giving the number of thermal photons at the frequency $\omega$ and $r$, $\varphi$ are the squeezing parameters. 

\begin{equation}
    H_{\Tilde{S}}= \hbar \sum_{j=1}^{2} \omega_{j} S_{j}^z + \hbar \sum_{m \neq n} ^2 \Omega_{mn} S_{m}^+ S_{n}^- ,
\end{equation}
where, 
\begin{align}
    \Omega_{mn} &= \frac{3}{4} \sqrt{\Gamma_m \Gamma_n} \bigg[-[1-(\hat{\mu}.\hat{r}_{mn})^2] \frac{\cos{(k_0 r_{mn})}}{k_0 r_{mn}} \nonumber \\
    +& [1-3(\hat{\mu}.\hat{r}_{mn})^2] 
    \times \left(\frac{\sin{(k_0 r_{mn})}}{(k_0 r_{mn})^2}  + \frac{\cos{(k_0 r_{mn})}}{(k_0 r_{mn})^3}\right) \bigg],
    \label{eq:atomic_frequencies}
\end{align}
Here, $\hat{\mu} =\hat{\mu}_1= \hat{\mu}_2 $ and $\hat{r}_{mn}$ are unit vectors along the atomic transition dipole moments and $r_{mn} = r_m - r_n$, respectively. Initially, the qubits are uncoupled but interact through the bath-mediated coupling represented by \(\Omega_{mn}\).
The ratio \(\frac{r_{mn}}{\lambda_0}\), which compares the interatomic distance to the resonant wavelength, determines the nature of the system's dynamics. When this ratio is large, we can express it as:  
\(k_0 r_{mn} \sim \frac{r_{mn}}{\lambda_0}\) (where, the ratio is on the order of or larger than unity), the system experiences independent decoherence, with each atom interacting with the environment individually. Conversely, when the ratio approaches zero;  
\(k_0 r_{mn} \sim \frac{r_{mn}}{\lambda_0}\), collective decoherence dominates. This regime arises when the qubits are sufficiently close to one another, allowing them to couple to the environment collectively, or when the bath's correlation length, characterized by the resonant wavelength \(\lambda_0\), is much larger than the separation between the qubits \(r_{mn}\).
The reduced density matrix Eq. (~\ref{eq:master_equation}), expressed in the two-qubit dressed state basis ~\cite{banerjee2010dynamics}, decompose into four independent blocks, denoted as A, B, C, and D and can be seen to produce X-type states. This block structure significantly simplifies the analysis by reducing the original set of fifteen coupled linear differential equations into four smaller, decoupled systems. At most, the solution requires addressing four coupled equations within any single block, offering a more efficient computational approach. The detailed solutions for these blocks are provided in ~\cite{banerjee2010dynamics}.

\begin{figure*}
    \centering
    \includegraphics[width=0.48\linewidth]{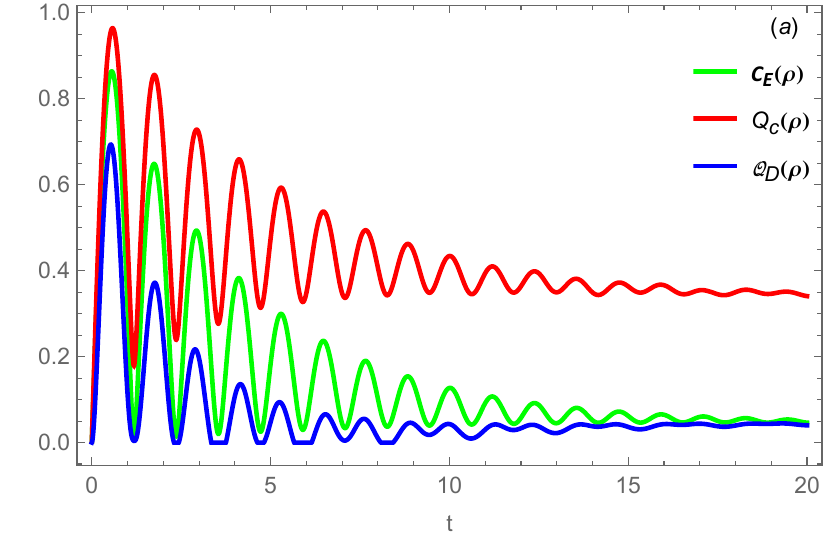}
    \includegraphics[width=0.48\linewidth]{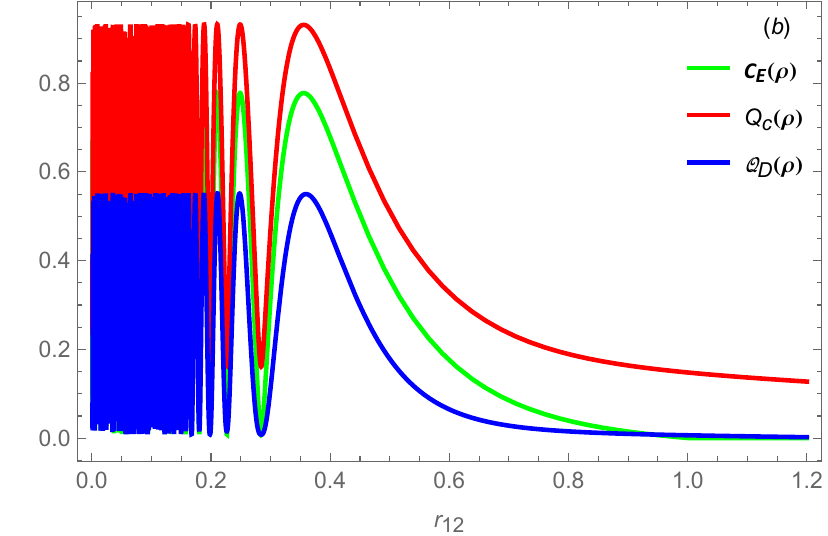}
    \caption{Variations of three different Quantum Correlations i.e. Concurrence \(C_{E}(\rho)\), Quantum Consonance \( Q_C(\rho)\) and Quantum Discord \(Q_D(\rho)\) are studied with a) time and b) interqubit distance (\(r_{12}\)). In a) the parameters are \(r_{12}\)= 0.2, $r= 0.35$, $T=1$. In b) the parameters are $t=1$, $r = 0.35$ and $T=1$.  }
    \label{fig:Comparision}
\end{figure*}
 
\section{Results and Discussions}
In this section, we present and analyze the results of modeling the dynamics of quantum correlations in a two-qubit system interacting with a squeezed thermal bath. The system is initially prepared in the separable state $|e \rangle | g\rangle$. \\
To begin, we investigate the decoherence effects induced by the squeezed thermal reservoir on the evolution of quantum correlations within the two-qubit system. The relative entropy of coherence precisely measures how much coherence is lost due to interactions with the squeezed thermal bath. It allows for a comparison between the initial and evolved states of the system, tracking the dynamics of coherence over time. In Fig. \ref{fig:relative_entropy}(a) we observe that in the collective regime, i.e., when the interqubit distance is relatively small, \(r_{12}=0.1\), at t=0, the coherence quantifier starts with a relatively high value we observe more fluctuations, and the system loses coherence with time. On the other hand, in the independent regime, when the interqubit distance is larger, \(r_{12}=1.1\), we observe no fluctuations. In Fig. \ref{fig:relative_entropy}(b), we observed that as we move from collective to independent regime, the oscillation frequency decreases and the system loses coherence.\\
\begin{figure*}
    \centering
    \includegraphics[width=0.48\linewidth]{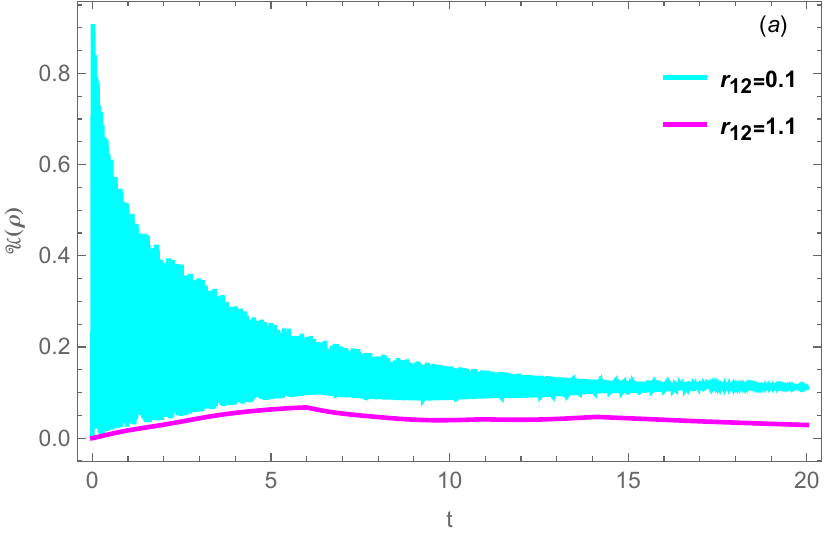}
    \includegraphics[width=0.48\linewidth]{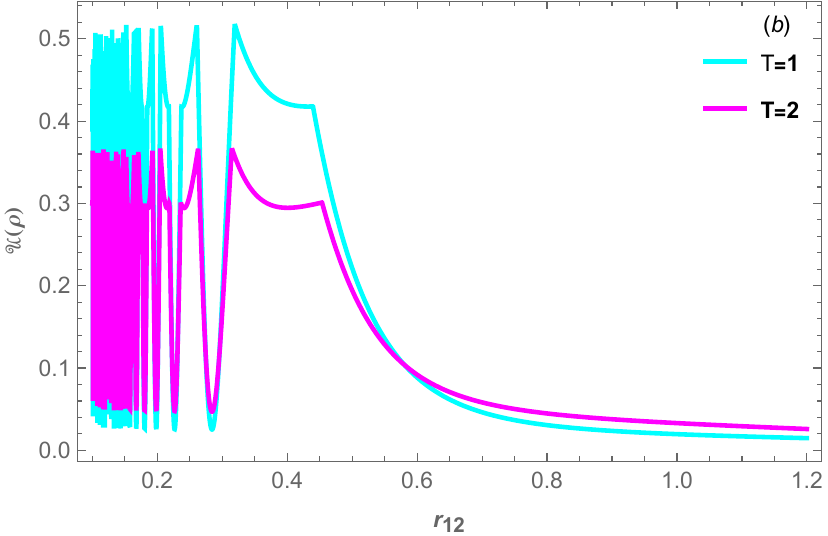}
    \caption{Variation of Local Quantum Uncertainty \(\mathcal{U}(\rho)\)  as a function of a) time (\(t\)) and b) inter-qubit distance (\(r_{12}\)) in a two-qubit system. In a) the time evolution is studied in the collective (\(r_{12}=0.1\)) and independent (\(r_{12}=1.1\)) regimes with parameters $r= 0.35$ and $T= 1$. In b) \(\mathcal{U}(\rho)\) variation with \(r_{12}\) is studied at two different temperatures $T= 1$ and $T=2$ with other parameters $r= 0.35$ and $t= 1$.}
    \label{fig:lqu}
\end{figure*}
It is well-established that quantum correlations extend beyond quantum entanglement. Notably, even separable states can exhibit non-classical behavior, as demonstrated in previous studies. This highlights the importance of exploring quantum correlations beyond entanglement. In this work, we delve into two additional forms of quantum correlations—quantum discord and quantum consonance—alongside entanglement to gain a more comprehensive understanding of the non-classical properties of quantum systems.
The evolution of quantum correlations in the separable initial state (\(|\psi\rangle = |e\rangle|g\rangle\)) is illustrated in Fig. \ref{fig:Comparision}. Figure  \ref{fig:Comparision}(a) reveals that concurrence, quantum consonance, and quantum discord exhibit oscillatory behavior in the collective regime, eventually stabilizing at a steady value. Figure  \ref{fig:Comparision}(b) highlights the correlation behavior at a fixed time (\(t\)) while varying the inter-qubit distance (\(r_{12}\)). As the system transitions from the collective to the independent regime, the oscillation frequency diminishes and stabilizes in the independent regime. Quantum consonance is the sum of quantum entanglement and some other non classical correlations, so it must always be larger than quantum entanglement ~\cite{cao2014dynamics}. Figures \ref{fig:Comparision} (a) and \ref{fig:Comparision} (b) demonstrate that this is the case. Quantum consonance is stronger than quantum entanglement, quantified here by concurrence, as demonstrated by the exhibited pattern. Figures \ref{fig:Comparision} (a) and \ref{fig:Comparision}(b) further confirm the earlier demonstration that quantum discord might be less than concurrence ~\cite{ali2010quantum,cao2014dynamics}. In Fig. \ref{fig:Comparision} (b) it is also observed that in the independent regime, concurrence and discord are almost zero but consonance is non-zero.

In Fig. \ref{fig:lqu}(a), LQU decays over time due to environmental decoherence, with the decay rate depending on the interqubit distance: shorter inter-qubit distances ($r_{12} = 0.1$) exhibit higher initial LQU and oscillatory behavior as it decays with time, while for larger inter qubit separations ($r_{12} = 1.1$) the oscillatory behavior is not observed. Figure \ref{fig:lqu}(b) reveals that LQU oscillates at till 0.4 ($r_{12} < 0.3$), and then monotonically decreases with increase in $r_{12}$.\\

\begin{figure*}
    \centering
    \includegraphics[width=0.48\linewidth]{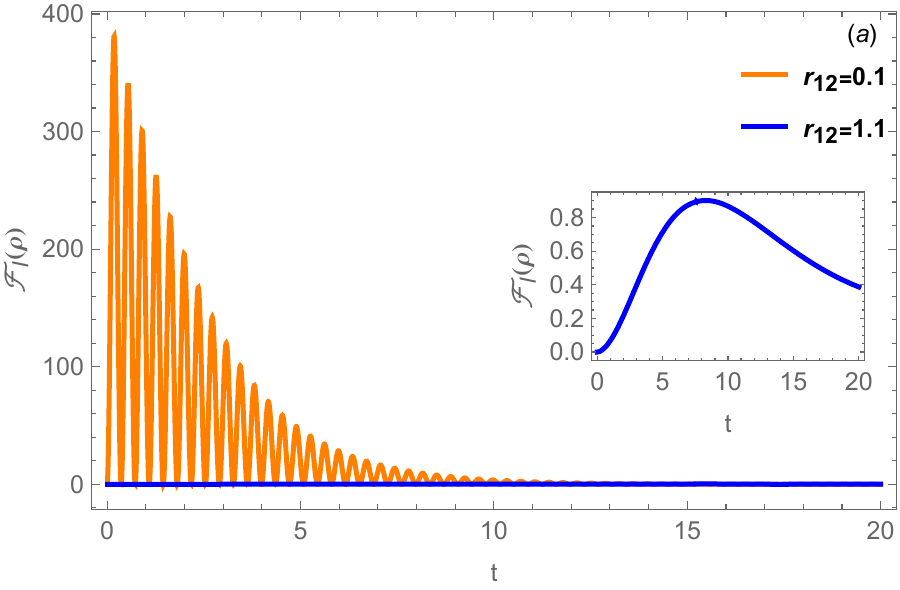}
    \includegraphics[width=0.48\linewidth]{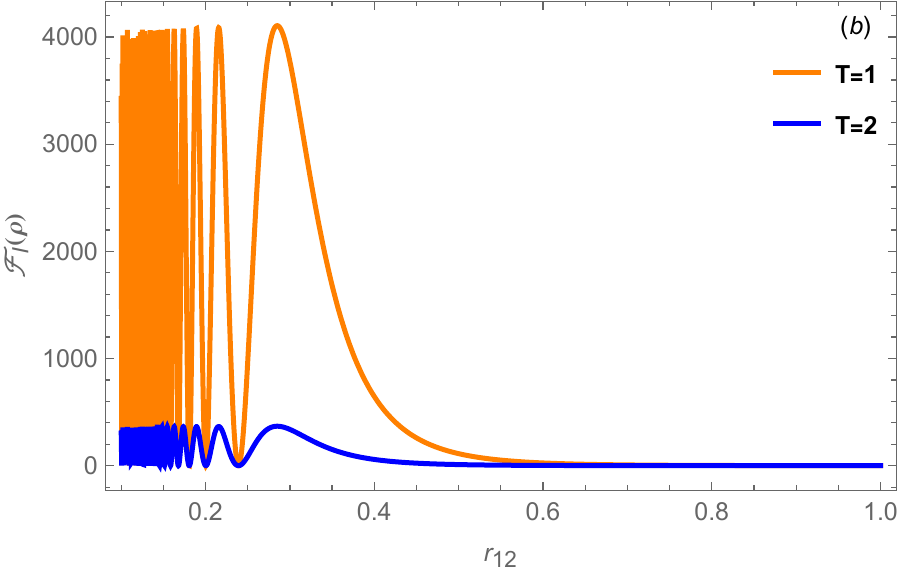}
    \caption{Variation of QFI ($\mathcal{F}_{\mathcal{I}} (\rho_{\vartheta}) $) is analyzed with a) time (t) b) interqubit distance (\(r_{12}\)). In a) the time evolution is studied in the collective (\(r_{12}=0.1\)) and independent (\(r_{12}=1.1\)) regimes with parameters $r= 0.35$ and $T=1$. The inset provides a zoomed in picture of the QFI in the independent regime. In b) QFI variation with \(r_{12}\) is studied at two different temperatures T= 1 and T= 2 with other parameters $r= 0.35$ and $t= 1$.}
    \label{fig:QFI}
\end{figure*}

Quantum Fisher Information (QFI) quantifies the information about a specific parameter of interest encoded in a quantum state. It represents the amount of physical information that the parameter carries within the quantum state, obtained through measurements and the process of estimating the parameter. In this work, we have considered $r_{12}$ as the parameter to be estimated \( \vartheta\). QFI is calculated using Eq. \eqref{fisher} and  the Fig. \ref{fig:QFI} illustrates the variation of QFI, \( \mathcal{F}_{\mathcal{I}} (\rho_{\vartheta})  \), w.r.t time (\( t \)) and interqubit distance (\( r_{12} \)). In Fig. \ref{fig:QFI}(a), for \( r_{12} = 0.1 \) (collective regime), QFI exhibits oscillatory behavior and decays over time. For \( r_{12} = 1.1 \) (independent regime), QFI remains nearly constant and significantly smaller. This is an indicator of higher sensitivity in the collective regime compared to the independent one. In Fig. \ref{fig:QFI}(b), the QFI variation with \( r_{12} \) exhibits fluctuations till  \( r_{12} \)= 0.2 to 0.4 after which QFI decreases to zero with further increase in \( r_{12} \) . It is also observed that this behavior is getting suppressed with an increase in tempertaure. Overall, QFI is more prominent in the collective regime, at lower temperatures, providing insights for optimizing parameter estimation in quantum metrology.

\begin{figure*}
    \centering
    \includegraphics[width=0.48\linewidth]{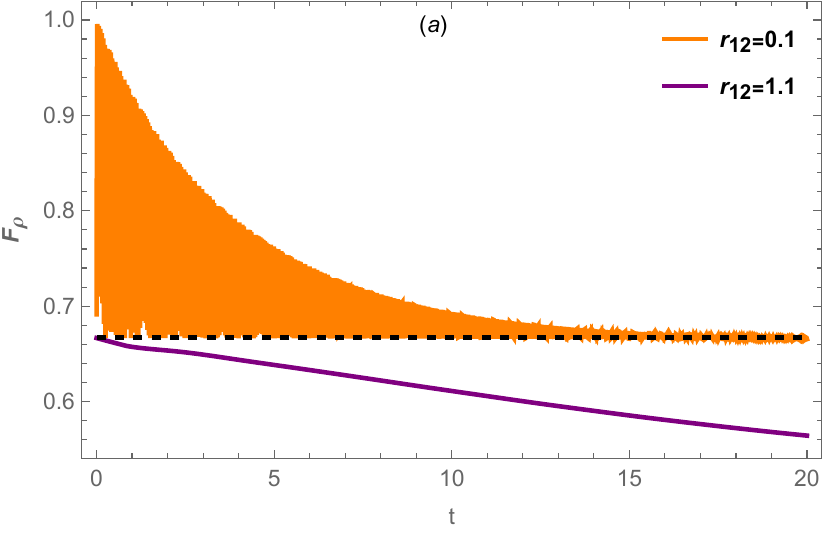}
    \includegraphics[width=0.48\linewidth]{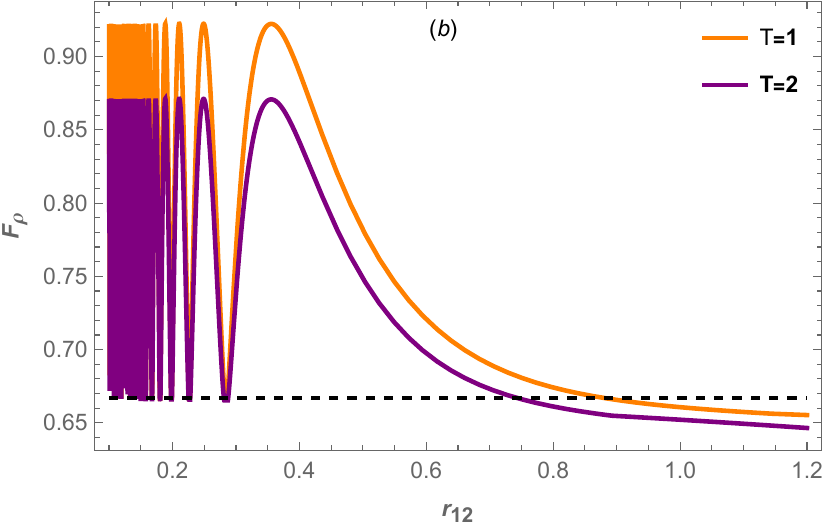}
    \caption{Variation of maximal teleportation fidelity \(\mathbf{F}_\rho\) with a) time (t) and b) interqubit distance (\(r_{12}\)) in a two-qubit system. In a) the time evolution is studied in the collective (\(r_{12}=0.1\)) and independent (\(r_{12}=1.1\)) regimes and parameters $r=0.35$ and $T=1$. In b) the plot shows the variation with interqubit distance (\(r_{12}\)) at two different values of temperature $T=1$ and $T=2$ with the squeezing parameter $r= 0.35$ and $t= 1$. The dashed line in both a) and b) shows the average teleportation fidelity value which is $2/3$.}
    \label{fig:fidelity}
\end{figure*}

To evaluate the efficiency of the quantum teleportation (QT) in mixed entangled states, we used two metrics: (i) maximal average fidelity and (ii) fidelity deviation \cite{ghosal2020fidelity,ghosal2021characterizing}. A two-qubit state is deemed useful for universal quantum teleportation if its maximal average fidelity exceeds \( 2/3 \). We assess the success by measuring the maximal teleportation fidelity \( F_\rho \) and fidelity deviation \( \Delta F_\rho \) using Eqs. \eqref{maxfidelity} and \eqref{fidelitydev}, respectively.
In Fig. \ref{fig:fidelity}, we observed the variations of maximal teleportation fidelity w.r.t time (t) and interqubit distance (\(r_{12}\)) in a two-qubit system. In Fig. \ref{fig:fidelity}(a), \( \mathbf{F}_\rho \) decreases over time, stabilizing near the classical fidelity threshold (\( \mathbf{F}_\rho = 2/3 \)), with higher fidelity in the collective regime (\( r_{12} = 0.1 \)) compared to the independent regime (\( r_{12} = 1.1 \)). In Fig. \ref{fig:fidelity}(b),  fidelity is seen to be suppressed with an increase in temperature. Fluctuations in fidelity are observed till  $r_{12} = 0.4$ and then it decreases and crosses the barrier of $2/3$ at \( r_{12} \approx 0.8 \). Overall, maximal teleportation fidelity is greater than $2/3$  for short interqubit distances and collective effects, indicating conditions necessary for effective quantum teleportation.\\

\begin{figure*}
    \centering
    \includegraphics[width=0.48\linewidth]{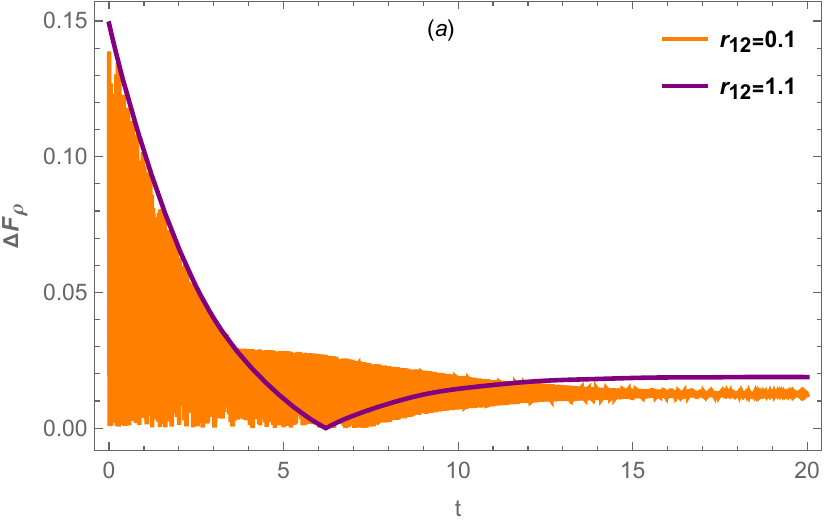}
    \includegraphics[width=0.48\linewidth]{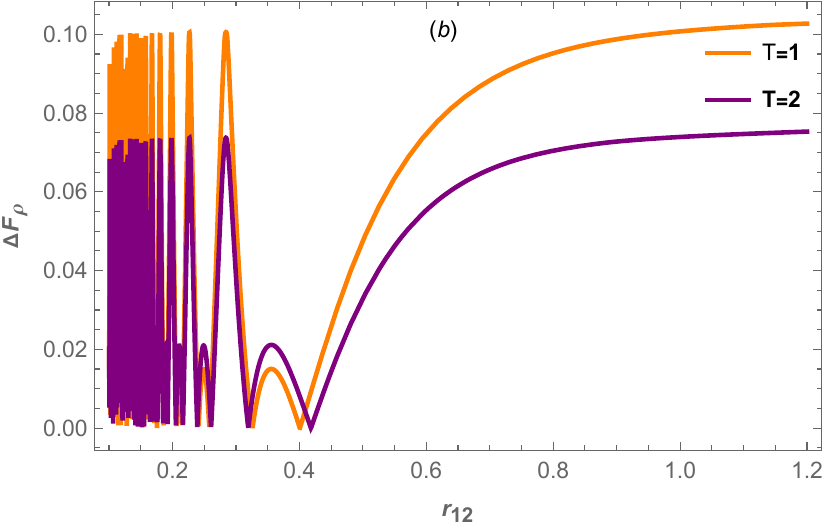}
    \caption{Variation of teleportation fidelity deviation \(\Delta \mathbf{F}_\rho\) with a) time (t) and b) interqubit distance (\(r_{12}\)) in a two-qubit system. In a) the time evolution is studied in the collective (\(r_{12}=0.1\)) and independent (\(r_{12}=1.1\)) regimes and parameters $r=0.35$ and $T=1$. In b) the plot shows the variation with interqubit distance (\(r_{12}\)) at two different values of temperature $T=1$ and $T=2$ with the squeezing parameter $r= 0.35$ and $t= 1$.}
    \label{fig:fidelity_deviation}
\end{figure*}

Figure \ref{fig:fidelity_deviation}, shows the variation of fidelity deviation \(\Delta \mathbf{F}_\rho\) with time and interqubit distance. In Fig. \ref{fig:fidelity_deviation}(a), \(\Delta \mathbf{F}_\rho\) decreases with time, with larger initial fluctuations in the collective regime (\( r_{12} = 0.1 \)), stabilizing near zero with time, while in the independent regime (\( r_{12} = 1.1 \))  \(\Delta \mathbf{F}_\rho\) decreases with time till t= 6 and then we see a subtle increase. As seen from Fig. \ref{fig:fidelity} (a), the value of \( \mathbf{F}_\rho \) in the independent regime is always less than 2/3 for the case considered here and hence, we can conclude that the QT is not useful in the independent regime. In Fig. \ref{fig:fidelity_deviation}(b), \(\Delta \mathbf{F}_\rho\) exhibits oscillatory behavior at short interqubit distances, showing a fluctuations till \(r_{12} \approx 0.4\) and then increases. This behavior is suppressed with an increase in temperature. To achieve a reliable and stable teleportation process, \(\Delta \mathbf{F}_\rho\) should be minimal along with \( \mathbf{F}_\rho\ > 2/3\), ensuring uniform fidelity for all input states. This can be confirmed for the collective decoherence regime from Figs. \ref{fig:fidelity_deviation}(a) and \ref{fig:fidelity_deviation} (b). \\
 
\section{Conclusion}
   Quantum correlations is a many-faceted entity. Here, we have examined the role of various quantifiers of non-classical correlations, specifically focusing on quantum consonance, concurrence, discord, local quantum uncertainty, and quantum coherence.  The quantum correlations were computed on a two-qubit model interacting with a squeezed thermal bath. This facilitated the study of the model in two different regimes: collective and independent. A generic feature observed was that the correlations observed have a fluctuating behavior in the collective regime. Comparing the non-classical correlations, consonance was seen to exceed the entanglement measure (concurrence), which aligns with the expected theoretical behavior.  \\
    In open systems, quantum correlations can be affected by decoherence or noise, and QFI provides a way to track how these factors influence the estimation precision. From our study of QFI, with the interqubit distance as the estimation parameter, it came out that the collective regime exhibits higher sensitivity. Further, the teleportation protocol was studied for this model from the perspective of maximal teleportation fidelity and fidelity deviation. It came out that QT is suitable in the collective regime.\\
   These factors are crucial in advancing our understanding of quantum information processing and the complex dynamics of open quantum systems.
\begin{acknowledgments}
Ramniwas Meena acknowledges financial support from CSIR-HRDG through the CSIR JRF/SRF fellowship ( File No. 09/1125(0015)2020-EMR-I).
\end{acknowledgments}
\bibliographystyle{apsrev}
\bibliography{main}
\end{document}